# Thermodynamic Limits of Spatial Resolution in Active Thermography

Peter Burgholzer

**Abstract** Thermal waves are caused by pure diffusion: their amplitude is decreased by more than a factor of 500 within a propagation distance of one wavelength. The diffusion equation, which describes the temperature as a function of space and time, is linear. For every linear equation the superposition principle is valid, which is known as Huygens principle for optical or mechanical wave fields. This limits the spatial resolution, like the Abbe diffraction limit in optics. The resolution is the minimal size of a structure which can be detected at a certain depth. If an embedded structure at a certain depth in a sample is suddenly heated, e.g. by eddy current or absorbed light, an image of the structure can be reconstructed from the measured temperature at the sample surface. To get the resolution the image reconstruction can be considered as the time reversal of the thermal wave. This inverse problem is ill-conditioned and therefore regularization methods have to be used taking additional assumptions like smoothness of the solutions into account. In the present work for the first time methods of non-equilibrium statistical physics are used to solve this inverse problem without the need of such additional assumptions and without the necessity to choose a regularization parameter. For reconstructing such an embedded structure by thermal waves the resolution turns out to be proportional to the depth and inversely proportional to the natural logarithm of the signal-to-noise ratio. This result could be derived from the diffusion equation by using a delta-source at a certain depth and setting the entropy production caused by thermal diffusion equal to the information loss. No specific model about the stochastic process of the fluctuations and about the distribution densities around the mean values was necessary to get this result.

___________________________

P. Burgholzer
Christian Doppler Laboratory for Photoacoustic Imaging and Laser Ultrasonics,
Research Center for Non Destructive Testing (RECENDT),
Altenberger Strasse 69, 4040 Linz, Austria.
E-Mail: peter.burgholzer@recendt.at

# 1  Introduction

In active thermography subsurface embedded structures are detected by heating the sample surface or the structures and measuring the time dependent surface temperature. The temperature evolution as a function of space and time is determined by the heat diffusion equation, and its solution can be described as a composition of plain thermal waves with different frequencies and wavenumbers [1]. The heat diffusion equation is a macroscopic mean-value-equation in the sense that in a microscopic picture the temperature is proportional to the mean value of the kinetic energy of the molecules. Stochastic thermodynamics describes the statistical distributions of the fluctuations around those mean values [2].

The detection and location of the embedded structures from the measured temperature evolution at the surface is an inverse problem. In the microscopic picture the movement of the molecules is invertible, as the kinetic equations are invariant to time reversal. The macroscopic mean-value-equation, for thermography the heat diffusion equation, is not invariant to time reversal any more. Recent results from non-equilibrium thermodynamics show that for macroscopic samples the mean entropy production is equal to the information loss by using the macroscopic mean-value-equation instead of the microscopic description (see section 2 and appendix). This loss of information for the macroscopic description is the physical reason that the inverse problem gets ill-posed and that subsurface structures cannot be detected any more if they are lying too deep under the surface.

Thermal waves are caused by pure diffusion: their amplitude is decreased by a factor of $e^{-2\pi} \approx 1/535$ within a propagation distance of one wavelength [1]. The diffusion equation, which describes the temperature as a function of space and time ("diffusion –wave fields" [3]), is linear. For every linear equation the superposition principle is valid, which is known as Huygens principle for optical or mechanical wave fields. This allows a composition of the solution of plane waves having different wavenumbers and similar to the wave equation also for the diffusion equation components with a higher wavenumber are attenuated more than those with a lower wavenumber. This limits the spatial resolution, like the Abbe diffraction limit in optics. The resolution is the minimal size of a structure which can be detected at a certain depth. If an embedded structure at a certain depth in a sample is suddenly heated, e.g. by eddy current or absorbed light, an image of the structure can be reconstructed from the measured temperature at the sample surface. One possibility for image reconstruction is a time reversal of the thermal waves. This inverse problem is ill-conditioned and therefore regularization methods have to be used taking additional assumptions like smoothness of the solutions into account. In the present work for the first time methods of non-equilibrium statistical physics are used to solve this inverse problem without the need of such additional assumptions and without the necessity to choose a regularization parameter.

Reconstructing the samples interior structure from the measured signals at the samples surface is a prominent example of an inverse problem. An inverse problem is a general framework that is used to convert observed measurements into information about a physical object or system. Inverse problems are typically ill-posed, as opposed to the well-posed problems more typical when modeling physical situations where the model parameters or material properties are known. Of the three conditions for a well-posed problem suggested by Jacques Hadamard [4] (existence, uniqueness, stability of the solution or solutions) the condition of stability is most often violated [5]. Even if a problem is well-posed, it may still be ill-conditioned, meaning that initially small errors can grow exponentially. An ill-conditioned problem is indicated by a large condition number. It needs to be re-formulated for numerical treatment. Typically this involves including additional assumptions, such as smoothness of solution [6]. This process is known as regularization, like truncated singular value decomposition (SVD) or Tikhonov regularization [7]. The choice of an adequate regularization parameter, which describes the trade-off between the original ill-conditioned problem and the additional assumptions (e.g. smoothness), is critical and has to be evaluated for every individual problem [6]. In this paper, we propose that by using the entropy production we get a physical background for choosing the regularization parameter for thermographic imaging and no additional assumption is necessary.

To simplify the following calculations two assumptions have been made. First, heat diffusion is assumed to take place only in one dimension. Physically this happens in layered structures, when the lateral dimension of these structures is big compared to their thickness. Second, the heat source is a thin layer which is instantaneous heated in a thick sample. Mathematically, this can be approximated by a Dirac delta function in infinite space in one dimension. For real thermography different boundary conditions on the sample surface have to be used. E.g. Mandelis has solved the heat diffusion equation in his book "Diffusion-Wave Fields" [3] in one, two, or three dimensions for different boundary conditions. The method demonstrated here to get a thermodynamic resolution limit for one dimensional heat diffusion in infinite space can be generalized to such boundary conditions and more dimensions, as discussed in section 4.

To get the spatial resolution usually a fine structure which is small compared to the resolution is imaged. Then the resolution is equal to the size of the imaged structure. For bigger structures the convolution of the structure with the blurring from image resolution gives the size of the imaged structure. Using the Dirac delta function as the initial temperature distribution ensures that the size of the imaged structure is just the resolution. Of course the evaluated resolution is valid for any initial temperature profile as the heat diffusion equation is a linear one. For the excitation in active thermography often a pulse is used, e.g. a light or eddy current pulse. The pulse duration is chosen to be short compared to the time needed for thermal diffusion along the samples structures. Mathematically this gives again a Dirac delta function, but in time. Signals for longer pulses can be calculated by a convolution integral using the temporal evolution of the generation pulse. Generalizations for other generation patterns, like a sinusoidal excitation for lock-in thermography, are discussed in section 4.

In the present work it should be shown that a general thermodynamic limit of the spatial resolution can be derived from a very recent result from stochastic thermodynamics [8], which is summarized in section 2 and is shortly reviewed in the Appendix: if a macroscopic system is kicked out of the equilibrium by a short but not necessarily small perturbation, such as a short laser pulse, for the following dissipative process back to equilibrium *the information loss about the kick magnitude is equal to the mean entropy production.* In active thermography the kick magnitude is no single number but the vector of the temperature in space just after the excitation pulse (= kick) or it can be the magnitude of all the Fourier components of the temperature wave, as used in section 3.

The information loss can be quantified by the Kullback-Leibler divergence (e.g. [9], also called relative entropy). The Kullback-Leibler divergence $D(f||g)$ is used in information theory for testing the hypothesis that the two distributions with density $f$ and $g$ are different [9] and is defined as

$$D(f||g) := \int ln\left(\frac{f(x)}{g(x)}\right) f(x) dx, \qquad (1)$$

where $ln$ is the natural logarithm. The Chernoff-Stein Lemma states that if $n$ data from $g$ are given, the probability of guessing incorrectly that the probability distribution for describing the data is $f$ is bounded by the type II error $\varepsilon = (\exp(-D(f||g)))^n$, for $n$ large [9]. In that sense $D(f||g)$ can describe some "distance" between the distribution densities $f$ and $g$.

The inverse problem of estimating the kick-magnitude from a measurement of an intermediate state a certain time after the kick is ill-conditioned. Just after the kick its magnitude can be estimated very well. A long time after the kick the state has nearly evolved back to equilibrium and all the information about the kick magnitude is lost. The information content at an intermediate state a time $t$ after the kick with a distribution density $p_t$ in comparison to the equilibrium distribution $p_{eq}$ is $D(p_t||p_{eq})$. For macroscopic systems $D$ is equal to the entropy production till time $t$ [8]. It has its maximum just after the kick, when no entropy has been produced yet. Then it decreases monotonously in time and gets zero in the limit of infinite time. But already at some earlier cut-off time $t_{cut}$ all the information about the kick magnitude is lost: according to the Chernoff-Stein Lemma for a fixed error $\varepsilon$ the distribution $p_t$ at time $t$ cannot be distinguished from the equilibrium distribution $p_{eq}$ if $D(p_t||p_{eq})$ gets smaller than $\ln(1/\varepsilon)/n$.

In active thermography it is assumed that the information about the spatial pattern from the interior structure to the surface of the sample is transferred by heat diffusion. The "thermal wave" can be represented as a superposition of wave trains having a certain wavenumber or frequency in Fourier $k$-space or $\omega$-space, respectively [10]. Instead of a cut-off time for the whole signal the decrease of the Kullback-Leibler divergence in Fourier space gives a criterion for a cut-off wavenumber or a cut-off frequency as an upper limit, where all the information about the Fourier-component is lost because it cannot be distinguished from equilibrium according the Chernoff-Stein Lemma. In the past we have modeled heat diffusion by a Gauss-Markov process in Fourier space and found a principle limit for the spatial resolution [10]. Using the information loss and entropy production for a kicked process it will be shown in section 2 (see also [8]) that the spatial resolution depends just on the macroscopic mean-value equations and is independent of the actual stochastic process, as long as the macroscopic equations describe the mean heat flow and therefore also the mean entropy production as the mean heat flow divided by the temperature. General limits of spatial resolution are derived in section 3 from the diffusion equation by using cut-off wavenumbers or frequencies from stochastic thermodynamics.

## 2 Information loss and Entropy Production for Kicked Processes

In active pulse thermography embedded structures are detected by heating the sample surface or the structures with a short excitation pulse, which "kicks" the sample out of its equilibrium. The kick is a sudden temperature rise due to e.g. optical absorption or electromagnetic induction at the sample structures. For the reconstruction very often the temperature distribution just after the kick is the "kick magnitude" which should be reconstructed, like in photothermal depth profiling of the first kind [11]. But the present theory is valid also for other applications: e.g., determining the thickness of sample sheets of opaque materials, like metals. When lightening the surface with a short light pulse the heat is absorbed in a very thin surface layer. It diffuses into the material, is "reflected" at the back plane (or at voids in the sample), and measured as a temperature change of the surface. For the thermodynamic resolution limit for such a case the depth a as given in (13) is the doubled thickness of the sheet as the thermal wave travels back and forth.

To derive the connection between information loss and entropy production a generalization of the second law and of Landauers principle for states arbitrarily far from equilibrium given recently by Hasegawa et al. ([12], [13]) and by Esposito and Van den Broeck in [14] is used. The main idea to deal with a non-equilibrium state $p_t$ is to perform a sudden transition from the known Hamiltonian $H$ with equilibrium state $p_{eq}$ to a new one $H^*$, such that the original non-equilibrium state becomes a canonical equilibrium with respect to $H^*$. The average amount of irreversible work for this quench turns out to be the Kullback-Leibler divergence $D(p_t||p_{eq})$ times $k_B T$, where $k_B$ is the Boltzmann constant and $T$ is the temperature of the system. The mean irreversible work for this quench is used either to change the entropy $\Delta S$ of the kicked system itself or to heat the surrounding by $\langle H \rangle_{p_t} - \langle H \rangle_{eq}$. For macroscopic samples it is shown in [8] and also shortly reviewed in the Appendix that the change in the system entropy $\Delta S$ can be neglected compared to $(\langle H \rangle_{p_t} - \langle H \rangle_{eq})/T$ as fluctuations in macroscopic systems are small compared to the changes in the mean value. Therefore the information $k_B D(p_t||p_{eq})$ about the non-equilibrium state any time $t$ after the kick can be approximated by the mean entropy production described by the diffusion equation.

In information theory $k_B D(p_t||p_{eq})$ can be identified as the amount of information that needs to be processed to switch from the known equilibrium distribution $p_{eq}$ (no kick) to the distribution $p_t$ a time $t$ after the kick [9]. If the logarithm to the base 2 is taken instead of the natural logarithm in (1), $D(p_t||p_{eq})$ measures the average number of bits needed to describe the kick magnitude, if a coding scheme is used based on the given distribution $p_{eq}$ rather than the "true" distribution $p_t$ [9]. The information theoretical interpretation of $D(p_t||p_{eq})$ according to the Chernoff-Stein Lemma states that $p_t$ cannot be distinguished from $p_{eq}$ at the cut-off time $t_{cut}$ if $D(p_{t_{cut}}||p_{eq})$ becomes smaller than $\ln(1/\varepsilon)/n$ (see Introduction). In the present work the error level $\varepsilon$ is chosen in a way that at the cut-off time the signal amplitude becomes less than the thermodynamic fluctuations (= noise level). In the next section this will be applied in the Fourier space to determine a cut-off wavenumber or a cut-off frequency.

## 3 Cut-off Wavenumber and Frequency in Fourier Space

In this section two different inverse problems for heat diffusion are presented. For the first inverse problem, the initial temperature profile $T(x,0)$ at a time $t=0$ should be reconstructed from the measured temperature $T(x,t_m)$ at one specific time $t_m > 0$. For the second inverse problem, the temperature $T(x_m,t)$ is measured at a specific point $x_m$, usually at the surface of the sample, for all times $t > 0$ and from that data the temperature $T(0,t)$ for the point $x = 0$ is reconstructed. The second inverse problem is more adequate for thermography, where the temperature can be measured only on the surface. Nevertheless the first inverse problem is very instructive and therefore it is also presented here.

In active thermography the macroscopic mean value equation for the temperature $T(x,t)$ as a function of time $t$ and space $x$ after the kick, which is a short heating pulse, is the diffusion equation

$$\left(\frac{\partial^2}{\partial x^2} - \frac{1}{\alpha}\frac{\partial}{\partial t}\right)T(x,t) = 0 \tag{2}$$

with $\alpha$ is the thermal diffusivity. For simplicity $x$ has only one dimension, but the same procedure can be used for a more dimensional space. The bilateral Fourier transform over space and its inverse are

$$\hat{T}(k,t) = \int_{-\infty}^{\infty} T(x,t)exp(ikx)dx$$
$$T(x,t) = \frac{1}{2\pi}\int_{-\infty}^{\infty} \hat{T}(k,t)exp(-ikx)dk \tag{3}$$

where $i = \sqrt{-1}$ and $k = 2\pi/\lambda$ is the wavenumber, with the wavelength $\lambda$. The wavelength quantifies the spatial resolution of $T(x,t)$ if the wavenumber in the integral of the Fourier transform is not taken till infinity but only to a limited wavenumber. This limitation of the wavenumber is caused by a lower cut-off-time for higher wavenumbers, where the Kullback-Leibler divergence $D$ gets too small, so that the wave trains with a higher wavenumber cannot be distinguished from equilibrium.

For the first inverse problem we could show in [7] that for adiabatic boundary conditions, where the sample is thermally isolated, the eigenfunctions are cosine-functions and in $k$-space (Fourier transform over space) the temperature evolution in time is a simple multiplication:

$$\hat{T}(k,t) = \hat{T}(k,0)exp(-k^2\alpha t) \tag{4}$$

The eigenvalues $exp(-k^2\alpha t)$ decrease with higher wavenumbers $k$ "exponentially" and for the inverse problem the multiplication with $exp(+k^2\alpha t)$ as a huge number makes the reconstruction unstable for higher wavenumbers. Therefore regularization was used, either truncated singular value decomposition (SVD) or Thikonov regularization. Results for both regularization methods were compared in [10]. The choice of an adequate regularization parameter, that is the cut-off value for the SVD or the trade-off parameter between the original ill-conditioned problem and the smoothness of the solution as an additional assumption for the Thikonov regularization, is critical. Li Voti et al. have described regularization by truncated SVD and genetic algorithms for photothermal depth profiling [11] where the influence of the number of used singular vectors on the reconstructed heat source profile is described. The L-curve method was used to find the optimum value of used singular vectors. Several groups, e.g. the group of A. Salazar [15], have investigated the reconstruction of thermal conducting depth-profiles from thermography data, e.g. by Thikonov regularization. This so-called photothermal inverse problem of the second kind is for small variations of the sample thermal properties as a function of depth mathematically the same problem as reconstructing the heat source profile [11]. In [7] the L-curve method was used to find the regularization parameter for Thikonov regularization and in [10] a certain stochastic process, the Ornstein-Uhlenbeck process, was used to derive the cut-off wavenumber for the SVD. In both cases the cut-off wavenumber $k_{cut}$ gave the same result as we derive in (7), but now without choosing a regularization parameter or using a specific stochastic process.

As initial condition at a time $t = 0$ the delta function $T(x, 0) = \delta(x)$ is taken, to be sure to get the imaging resolution and not a convolution with the initial structure (see Introduction), which results in a constant Fourier transform $\hat{T}(k, t = 0) = 1$. All wavenumbers till infinity are present, which gives the best spatial resolution. $\lambda$ going to zero means that even peaks which are only separated by a very small distance can be still reconstructed as two separate peaks. After a certain time $t$ one gets from thermodynamics the mean entropy in $k$-space (e.g. [10], [2]) proportional to $\hat{T}(k, t)^2$, which shows an exponential decay in time with $exp(-2k^2 \alpha t)$, see (4):

$$\Delta S_k(t) = \frac{1}{2} k_B SNR^2 \, exp(-2k^2 \alpha t) \tag{5}$$

with the signal-to-noise ratio $SNR$. The wavenumber $k_{cut}$ is determined by using the Chernoff-Stein Lemma:

$$\Delta S_{k_{cut}}(t) \approx k_B D_{k_{cut}}(p_t || p_{eq}) \approx k_B \frac{1}{n} ln\left(\frac{1}{\varepsilon}\right) \tag{6}$$

If error $\varepsilon$ is set to $1/\sqrt{e}$ and $n = 1$ for one measured temperature one gets that the cut-off wavenumber $k_{cut}$ is just the wavenumber for which at a time $t$ the signal $\hat{T}(k_{cut}, t)$ gets less than the noise-level (Fig. 1). Then

$$k_{cut} = \sqrt{\frac{\ln SNR}{\alpha t}} \tag{7}$$

The reconstructed signal $T_r(x, t)$ is the Fourier transform of a rectangular function, which gives a sinc-function (see Fig. 1):

$$T_r(x, t) = \frac{1}{2\pi} \int_{-k_{cut}}^{k_{cut}} exp(-ikx) dk = \frac{1}{\pi} \frac{\sin(k_{cut}(t) x)}{x}. \tag{8}$$

The resolution $\delta_r(t)$ is the "width" of the reconstructed signal and is taken as the distance between the zero points of the reconstructed signal $T_r(x,t)$ – to be on the save side – which is the wavelength corresponding to the wavenumber $k_{cut}$:

$$\delta_r(t) = \frac{2\pi}{k_{cut}(t)} = 2\pi\sqrt{\frac{\alpha t}{\ln SNR}}. \qquad (9)$$

This is the same result as derived in [10] where we assumed that the thermal diffusion in $k$-space is an Ornstein-Uhlenbeck process, but in the derivation above the knowledge of the specific stochastic process is not necessary. The resolution given in (9) depends only on the diffusion equation (2) as the equation for the mean value of the stochastic process.

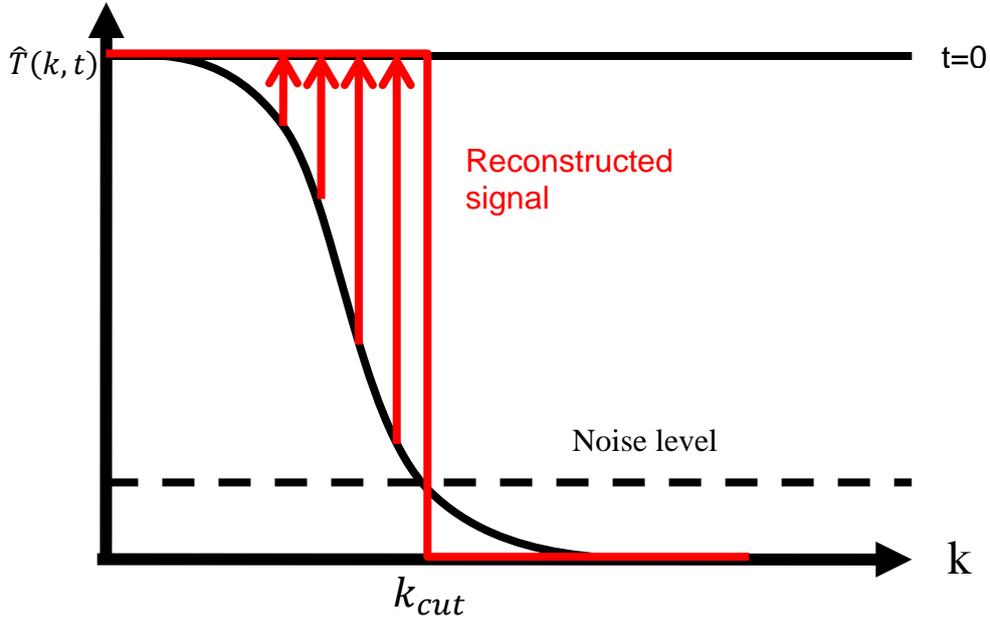

**Fig. 1** Temperature in $k$-space: $\hat{T}(k,t)$ shows an exponential decay in time with $exp(-k^2\alpha t)$. At $k_{cut}$ it goes below the noise level. For the reconstruction only the wavenumbers less than $k_{cut}$ are taken.

Usually in thermography the temperature is not measured on the whole sample $x$ at a certain time $t$, but at a certain $x$ – which is usually the sample surface – the temperature is measured at several times $t$. Instead of a Fourier transform (3) to $k$-space a bilateral Fourier transform in $\omega$-space is performed:

$$\tilde{T}(x,\omega) = \int_{-\infty}^{\infty} T(x,t)exp(-i\omega t)dt$$
$$T(x,t) = \frac{1}{2\pi}\int_{-\infty}^{\infty} \tilde{T}(x,\omega)exp(i\omega t)d\omega. \qquad (10)$$

Now for a certain depth $a$ from the Fourier transform of the diffusion equation, which is the Helmholtz equation for $\tilde{T}(x,\omega)$, a cut-off frequency $\omega_{cut}$ is determined by using the Chernoff-Stein Lemma analog to (6):

$$\omega_{cut} = 2\alpha \left(\frac{\ln SNR}{a}\right)^2 \quad (11)$$

This is consistent to a damping with the thermal diffusion length $\mu = \sqrt{2\alpha/\omega_{cut}}$, where the signal is damped by a factor of $\exp(-a/\mu)$ to the noise level during propagation along the length $a$ (e.g. [1] or [3]).

The reconstructed signal $T_r(a)$ at $x = 0$ is:

$$T_r(a) = \frac{1}{\pi}\int_0^{\omega_{cut}} \frac{1}{\sqrt{\omega\alpha}} \cos\left(a\sqrt{\frac{\omega}{2\alpha}}+\frac{\pi}{4}\right)\exp\left(-a\sqrt{\frac{\omega}{2\alpha}}\right)d\omega$$
$$= \frac{2}{\pi}\frac{\sin(k(a)a)}{a}\exp(-(k(a)a)) \text{ with } k(a) = \sqrt{\frac{\omega_{cut}}{2\alpha}} = \ln(SNR)/a \quad (12)$$

Like in (9) the resolution $\delta_r(a)$ is taken as the distance between the zero points of the reconstructed signal $T_r(a)$, which is the wavelength corresponding to the wavenumber $k(a)$:

$$\delta_r(a) = \frac{2\pi}{k(a)} = 2\pi \frac{a}{\ln SNR}, \quad (13)$$

which is proportional to the depth $a$ and independent from the thermal diffusivity $\alpha$. The resolution is proportional to the thermal diffusion length at the cut-off frequency.

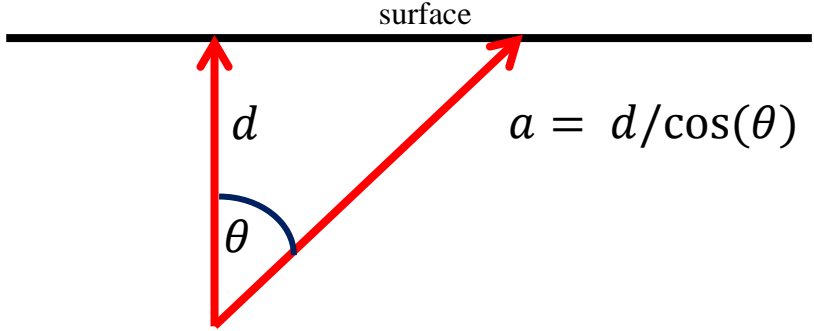

**Fig. 2** Thermal waves which do not go directly to the surface but at an angle $\theta$ have a longer path $a$ and therefore their minimal wavelength which can be detected on the surface is increased by a factor $1/cos(\theta)$.

This is not only the limit for the depth resolution, but also for the lateral resolution as can be estimated by using the formula for the Abbe diffraction limit $\delta_{Abbe} = \lambda/(2\sin(\theta))$. The minimal wavelength $\lambda = 2\pi\sqrt{2\alpha/\omega_{cut}}$ of the thermal wave as a certain point at the surface is now a function of the angle $\theta$, as the path $d$ at a certain angle $\theta$ is stretched to $a = d/\cos(\theta)$ (Fig. 2). This reduces the cut-off frequency $\omega_{cut}$ by a factor of $cos(\theta)^2$, which gives for the Abbe limit:

$$\delta_{Abbe}(\theta) = \frac{\lambda}{2\sin(\theta)} = \frac{\pi a}{\ln(SNR)\sin(\theta)} = \frac{\pi d}{\ln(SNR)\sin(\theta)\cos(\theta)}. \quad (14)$$

The term $\sin(\theta)\cos(\theta)$ is maximal for a value of $\theta = 45°$, where its value is ½:

$$\delta_{Abbe}(\theta = 45°) = 2\pi \frac{d}{\ln SNR} \qquad (15)$$

This expression for the lateral resolution is the same as given in (13) for the depth resolution. The detection circle at the surface has a radius of $d$. At a higher distance from the center ($\theta > 45°$), the resolution is cut because of diffusion. Compared to optics the influence of the minimal wavelength varying with $\theta$ on the Abbe resolution limit is not fully clear. This might result in a better resolution as estimated in (15); see also the discussion in section 4.

## 4  Discussion, Conclusions and Outlook

Eq. (13) and (15) are the main result of this work. For thermographic depth profiling the axial and the lateral resolution are proportional to the depth and inversely proportional to the natural logarithm of the signal-to-noise ratio. The resolution does not depend on the thermal diffusivity $\alpha$. This result could be derived from the diffusion equation by using a delta-source at a certain depth and setting the entropy production caused by thermal diffusion equal to the information loss. The delta-source in space as a point source is used to be sure to have a smaller structure than the blurring from image resolution. The derived resolution is valid for any initial temperature profile as the heat diffusion equation is a linear one. No specific model about the stochastic process of the fluctuations and about the distribution densities around the mean values was necessary to get this result. In earlier work [10] we derived eq. (9) by assuming that the thermal diffusion in $k$-space is a special Gauss-Markov process (Ornstein-Uhlenbeck process). The same result in $k$-space was derived in [7] by choosing for the Thikonov regularization the regularization parameter equal to the inverse signal-to-noise ratio, which was justified by the "L-curve" method. In the present publication e.g. the cut-off wavenumber $k_{cut}$ in (7) was derived by using a recently gained thermodynamic result: that the mean entropy production is equal to the information loss. In comparison to previous publications on photothermal depth profiling e.g. by Li Voti [11], the groups of Salazar [15], or Majaron et al. [16], where the L-curve or the Morozov discrepancy principle was used to determine the regularization parameters we could derive a principle limit from thermodynamics and no additional assumption for regularization or of a specific model for the stochastic process was necessary.

This result is consistent with work about interference of thermal waves [17]. The wavenumber for a thermal wave with frequency $\omega$ is $\sqrt{\omega/2\alpha}$, which is the inverse thermal diffusion length (e. g. [1]). Using the frequency $\omega_{cut}$ from (11) the wavelength corresponding to this wavenumber is just the resolution in (13). Also experimental results indicate the linear relation between depth and resolution, e.g. in [16]. In the derivation of eq. (15), using the formula for the optical Abbe diffraction limit, the minimal wavelength varies with the aperture-angle. This is different to optics, where the whole aperture has the same wavelength. Therefore the resolution given in (15) might be an upper limit. This has to be verified by additional two- or three-dimensional models and simulations.

For future work instead of a short kick also other excitation patterns can be considered to evaluate the resolution limits. G. Busse used a phase angle measurement with a sinusoidal excitation in lock-in thermography to get a better resolution in depth [18]. Sreekumar and Mandelis proposed a chirped excitation pattern like in radar technology to get a better spatial resolution at a certain depth [19]. Using the mean entropy production it should be possible to give also thermodynamic resolution limits for those excitation patterns and compare them to resolution limits for single short pulse excitation. Different and more realistic boundary conditions for the surface should be implemented (e.g. third kind in [3]) to be comparable to experimental results. First results for such a third kind boundary condition are given in [7], where we have heated a metal foil embedded in epoxy resin by a short eddy current pulse. Instead of the $k$-space, where the eigenfunctions are e.g. cosine-functions, new base functions $f_n$ were used (mathematically the same functions as in quantum physics for the finite potential well, e.g. Griffiths [20]).

**Appendix**

To show the equality of information loss and entropy production for the kicked process we follow the derivation of Kawai et al. [21] and Gomez-Marin et al. [22] in spirit to that of the Jarzynski [23] and Crooks [24,25] equalities, but instead of varying a control parameter λ from an initial value to a final value along a given protocol as in [21] we assume to start from a canonical equilibrium state at a temperature $T$ and "kick" it at time $t = 0$. We consider a Hamiltonian $H(x)$. $x$ is a point in phase space, where $x = (q,p)$ represents the set of position and momentum coordinates. Before the kick the equilibrium probability distribution to observe the state $x$ is given by a Boltzmann distribution $p_{eq}(x) = \exp(-\beta H(x))/Z$. $Z$ is the normalization factor (partition function) and $\beta := 1/(k_B T)$. The distribution density just after the short kick is $p_{kick}(x) = p_{eq}(x - x_0)$. The applied work $W$ for a kick $x_0$ is $H(x + x_0) - H(x)$ for a phase point $x$ at $t=0$. With

$$e^{\beta W} = \frac{e^{-\beta H(x)}}{e^{-\beta H(x+ x_0)}} = \frac{p_{eq}(x)}{p_{eq}(x + x_0)} \tag{16}$$

one gets by averaging the logarithm $ln$ of (16) and substituting $x' = x + x_0$:

$$\beta \langle W \rangle_{eq} = \int ln\left(\frac{p_{eq}(x)}{p_{eq}(x + x_0)}\right) p_{eq}(x) dx$$
$$= \int ln\left(\frac{p_{eq}(x' - x_0)}{p_{eq}(x')}\right) p_{eq}(x' - x_0) dx' \tag{17}$$

or multiplied by $k_B$ and using the definition of the Kullback-Leibler divergence (1):

$$information\ loss = k_B D(p_{kick}||p_{eq}) = \frac{\langle W \rangle_{eq}}{T} = \frac{1}{T}(\langle H \rangle_{kick} - \langle H \rangle_{eq})$$
$$= mean\ entropy\ production \tag{18}$$

In this equation the distribution density $p_{kick}$ is a shifted equilibrium density corresponding to a Hamiltonian $H(x - x_0)$. The partition function $Z$ is the same for $p_{kick}$ and $p_{eq}$, therefore $\Delta F$ is zero. More applied work $\langle W \rangle$ in (18(18) means that the distribution $p_{kick}$ just after the kick is "more distant" from the equilibrium distribution $p_{eq}$. The equilibrium is the state where all the information about the kick is lost and all the applied work has been dissipated.

A certain time $t > 0$ after the kick only a part of the applied work has been dissipated and not all the information about the kick magnitude has been lost. To describe a state at a time $t$, which is usually not an equilibrium state, it is not necessary to know $p_t$ for all microscopic variables, but $x$ as a set of reduced variables which captures the information on the work is sufficient [8]. In Fig. 3 the forward process is illustrated and the distributions are sketched. We propose that (18) can be written in a time-dependent form for all the intermediate non-equilibrium states after the kick with a distribution density $p_t$ in a good approximation.

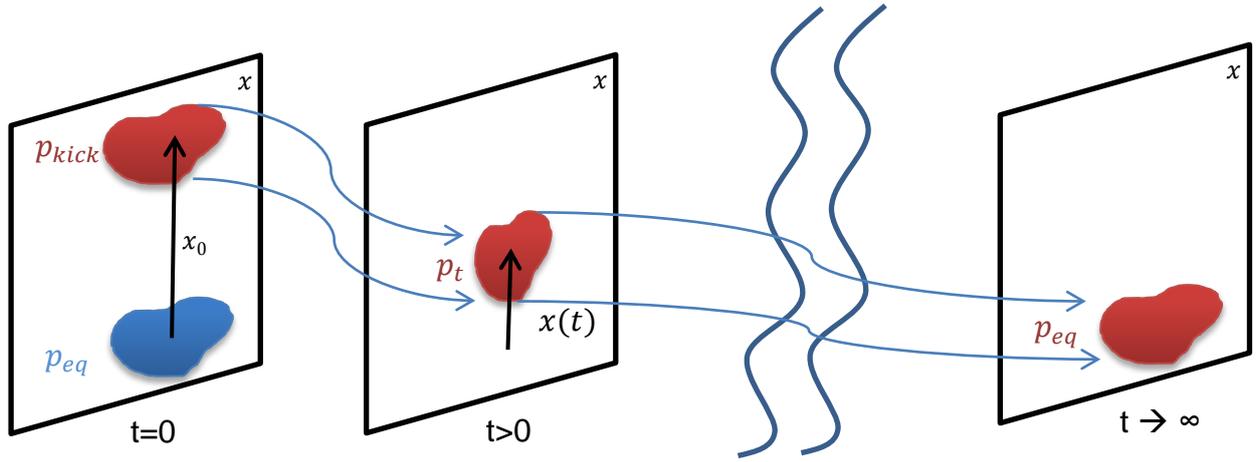

**Fig. 3** Illustration of the forward process: a system in equilibrium state $p_{eq}$ with mean value at $x = 0$ is kicked at a time $t=0$ with magnitude $x_0$ to a state $p_{kick}$ far from equilibrium, followed by a dissipative process back to equilibrium. $x$ is a set of reduced variables which captures the information on the work. The arrows connecting $p_{kick}$ at time $t=0$, $p_t$ at $t > 0$, and $p_{eq}$ at $t \to \infty$ indicate the tube of trajectories, which is "thin" for macroscopic systems as deviations from the mean values $x(t)$ are small.

Using the definition of the Kullback-Leibler divergence $D$ and $p_{eq}(x) = \exp(-\beta H(x))/Z$ one gets:

$$\text{information loss} = k_B \Delta D := k_B(D(p_{kick}||p_{eq}) - D(p_t||p_{eq}))$$
$$= \Delta S + \frac{1}{T}(\langle H \rangle_{kick} - \langle H \rangle_{p_t}) = \text{mean entropy production.} \quad (19)$$

The entropy production is the total entropy change $\Delta S$ minus the entropy flow, which is the negative of the dissipated heat $\langle H \rangle_{kick} - \langle H \rangle_{p_t}$ divided by the temperature. The total entropy change $\Delta S \equiv S(t) - S_{kick}$ is the difference in the Shannon entropy of $p_t$ and $p_{kick}$:

$$S(t) := -k_B \int p_t(x) \ln(p_t(x)) \, dx. \quad (20)$$

As $p_{kick}$ is only a "shifted" equilibrium distribution $p_{eq}$, the two distributions have the same entropy: $S_{kick} = S_{eq}$. Equation (19) can be also deduced directly from equation (12) given by Esposito and Van den Broeck in [14] by taking a time-invariant equilibrium distribution. Then $W_{irr} = 0$, which is the sum of the entropy production $\Delta_i S = \Delta S - Q/T$ and the information $\Delta I = -k_B \Delta D$. $Q$ is the heat coming from the heat bath, in our case the negative of the dissipated heat $\langle H \rangle_{kick} - \langle H \rangle_{p_t}$.

Equation (19) describes that the information $\Delta I$ about the kick magnitude decreases and the information loss increases during the evolution of time and is equal to the mean dissipated work divided by the temperature plus the entropy change $\Delta S$. After a long time $t$ the distribution $p_t$ converges to the equilibrium distribution $p_{eq}$ and from equation (19) one gets (18) using $\Delta S = S_{eq} - S_{kick} = 0$. This is also true in the linear regime near equilibrium as the shape of the distribution $p_t$ and therefore $S(t)$ does not change and is equal to $S_{eq}$. But also far from equilibrium in a good approximation for all the intermediate states $\Delta S \ll$ $1/T$ $(\langle H \rangle_{kick} - \langle H \rangle_{p_t})$, as for a macroscopic system fluctuations are small compared to the mean value (see Fig. 3 showing a "thin" tube of trajectories). Then the distribution $p_t$ has nearly no overlap with $p_{eq}$ and one gets:

$$D(p_t||p_{eq}) = \int p_t \ln p_t \, dx - \int p_t \ln p_{eq} \, dx \approx - \int p_t \ln p_{eq} \, dx = \beta \langle H \rangle_{p_t} + \ln Z. \quad (21)$$

The entropy term $p_t \ln p_t$ in (21) can be neglected because for all regions in the phase space where $p_t$ is different from zero and which contribute to the integral, $p_{eq}$ is nearly zero and $\ln p_t$ can be neglected compared to $\ln p_{eq}$. The same approximation in (21) is valid for $p_t = p_{kick}$ and therefore the total entropy change $\Delta S$ in (19) can be neglected:

$$k_B(D(p_{kick}||p_{eq}) - D(p_t||p_{eq})) \approx \frac{1}{T}(\langle H \rangle_{kick} - \langle H \rangle_{p_t}). \quad (22)$$

Subtracting (22) from (18) one gets for the information $I(t)$ about the kick magnitude:

$$I(t) = k_B D(p_t||p_{eq}) \approx \frac{1}{T}(\langle H \rangle_{p_t} - \langle H \rangle_{eq}) \approx \frac{1}{T}\Big(H(x(t)) - H(x=0)\Big), \quad (23)$$

$I(t)$ can be identified as the amount of information that needs to be processed to switch from the known equilibrium distribution $p_{eq}$ (no kick) to the distribution $p_t$ a time $t$ after the kick [9]. All the information about the kick can be approximated by the mean work, which has not been dissipated yet, divided by the temperature. This is the entropy production according to heat diffusion. After a long time all the energy has been dissipated and no information about the kick magnitude is available. The second approximation in (23) uses that for a macroscopic system the fluctuations are small and the mean of the Hamiltonian is approximately the Hamiltonian of the mean value $x(t)$ (Fig. 3).

**Acknowledgments** This work has been supported by the Austrian Science Fund (FWF), project number S10503-N20, by the Christian Doppler Research Association, the Federal Ministry of Economy, Family and Youth, the European Regional Development Fund (EFRE) in the framework of the EU-program Regio 13, and the federal state of Upper Austria.